\documentclass[prb,preprint,showpacs,preprintnumbers,amsmath,amssymb]{revtex4}

\usepackage{graphicx} 
\usepackage{dcolumn} 
\usepackage{bm}
\usepackage{txfonts}
\usepackage{mathrsfs}
\usepackage{comment}

\def\vk{{\bf k}}

\def\bra{\langle}
\def\ket{\rangle}
\newcommand{\eq}[1]{Eq.~(\ref{#1})}
\newcommand{\fig}[1]{Fig.~\ref{#1}}
\newcommand{\be}{\begin{equation}}
\newcommand{\ee}{\end{equation}}
\newcommand{\bea}{\begin{eqnarray}}

\newcommand{\eea}{\end{eqnarray}}

\begin{document}

\title{Raman scattering near a $\boldsymbol{d}$-wave Pomeranchuk instability} 

\author{Hiroyuki Yamase$^{1}$ and Roland Zeyher$^{2}$}
\affiliation{$^{1}$National Institute for Materials Science, Tsukuba 305-0047, Japan\\
$^{2}$Max-Planck-Institute for Solid State Research, Heisenbergstr.~1, D-70569 Stuttgart, Germany} 

\date{\today}

\begin{abstract}
Motivated by recent transport and neutron scattering experiments suggesting an 
orientational symmetry
breaking in underdoped cuprates we present a theoretical study of Raman scattering 
near a $d$-wave Pomeranchuk instability (PI). 
The $d$-wave component of Raman scattering  
from electrons and phonons allows to study directly order
parameter fluctuations associated with the PI.
Approaching the PI from the normal state by lowering the temperature
a central peak emerges both in electronic and, as an additional low-frequency
feature, in phononic scattering. Approaching the PI in the superconducting state
at low temperature by decreasing the doping concentration the central peak is replaced by
a soft mode with strongly decreasing width and energy and increasing 
spectral weight. These predicted low-energy features in Raman scattering could
confirm in a rather direct way the presence of a PI in
high-temperature cuprate superconductors and in Sr$_3$Ru$_2$O$_7$. 
\end{abstract}



\maketitle

\section{Introduction}
In condensed matter electrons move through a crystal lattice whose symmetry 
is characterized by a point group. 
The electronic band structure usually has the same symmetry 
as the lattice and so does the Fermi surface. 
However, it was shown that the symmetry of the  Fermi surface can be broken 
spontaneously by electron-electron correlations in the two-dimensional 
$t$-$J$\cite{yamase00,miyanaga06,edegger06} and 
Hubbard\cite{metzner00,valenzuela01} models leading to spontaneous 
Fermi surface deformations characterized by a $d$-wave symmetry [$d$-wave 
Fermi surface deformations ($d$FSD)]. 
This instability is frequently referred to as 
a $d$-wave Pomeranchuk instability, which is characterized by the violation of
the stability criteria for isotropic Fermi liquids derived by Pomeranchuk.\cite{pomeranchuk58} 
However, it should be noted that the $d$FSD state can also be realized 
not only in strongly correlated electron systems such as those described by 
the $t$-$J$ model\cite{yamase00,miyanaga06,edegger06} but also 
without a breaking of  
Pomeranchuk's stability criterion in systems where the transition can become 
of first order at low temperatures.\cite{kee0304,yamase05} 
The $d$FSD state breaks only the orientational 
symmetry, that is,  
its instability is driven by zero momentum charge-density 
fluctuations with internal $d$-wave symmetry 
and leads to an electronic nematic state. 
As originally introduced in Ref.~\onlinecite{kivelson98}, 
an electronic nematic state can also be realized by invoking charge stripes. 
In a first step both orientational and translational symmetry are broken 
by condensing the electrons into a charge stripe state 
characterized by a set of large wave vectors which break the orientational symmetry. 
In a second step the stripes melt restoring 
the translational but not the orientational symmetry. 
In the following we restrict
ourselves to the case where the $d$FSD leads directly to an 
electronic nematic state without showing first an instability towards stripes.

The double-layer strontium ruthenate 
Sr$_3$Ru$_2$O$_7$ (Sr327) has attracted much attention as a compound 
likely exhibiting a $d$FSD instability.\cite{grigera04,borzi07,rost09}   
Compelling, but indirect, evidence for this comes from the observation of a strong 
$xy$ anisotropy of the resistivity which is present only in the ordered phase.\cite{borzi07}
Angle-resolved photoemission spectroscopy (ARPES)\cite{tamai08} 
and de Hass-van Alphen\cite{perry04,borzi04,mercure09}  
measurements could detect Fermi surface deformations directly, 
but convincing experimental evidence for their existence has not been obtained yet. 
Theoretically many properties have been successfully interpreted in terms of
a $d$FSD instability, for instance,
the metamagnetic transition,\cite{kee05} 
the enhancement of the residual resistivity,\cite{doh07} 
the phase diagram and various thermodynamic quantities,\cite{yamase07b}  
universal numbers,\cite{yamase07c} 
the bilayer effect,\cite{puetter07,yamase09b}  
suppression of a critical temperature due to impurities,\cite{ho08} 
the spin-orbit effect,\cite{fischer10} and orbital degree of freedom.\cite{puetter10} 
Theoretical predictions based on the $d$FSD instability were also made for
the pattern of Friedel oscillation around an impurity,\cite{doh07b} 
the attenuation of ultrasound waves,\cite{adachi09}  
and the singular behavior of the uniform magnetic susceptibility at the $d$FSD instability.\cite{yamase10} 

In the case of the high-temperature superconductors 
YBa$_2$Cu$_3$O$_y$ (YBCO$_{y}$), 
the dynamical in-plane magnetic susceptibility is strongly 
anisotropic, both for 
slightly underdoped (YBCO$_{6.6}$)\cite{hinkov04,hinkov07} 
and optimally doped (YBCO$_{6.85}$)\cite{hinkov04}  compounds. 
The anisotropy increases with decreasing doping and is most pronounced
around the onset temperature of superconductivity or of the pseudogap,
whereas it is suppressed in the superconducting state. 
It was shown theoretically that these features can be well understood in 
terms of the competition of the singlet pairing formation and $d$FSD 
correlations.\cite{yamase06} 
In the strongly underdoped region (YBCO$_{6.45}$) neutron scattering experiments
revealed a qualitatively different feature of the anisotropy.\cite{hinkov08} 
The in-plane anisotropy of the magnetic excitation spectrum  
increases monotonically below 150 K, saturates below 50 K, but is 
not suppressed below $T_c=35$ K. 
Moreover, the low-energy spectral weight does not decrease below $T_c$
but is rather enhanced. These peculiar phenomena can be interpreted as 
i) a quantum phase transition to the $d$FSD state deep inside 
the superconducting state\cite{kim08,huh08}  
or ii) a substantial suppression of singlet pairing due to the competition 
with increasing $d$FSD correlations in the strongly 
underdoped region.\cite{yamase09} 

Quite recently the measurement of the Nernst coefficient 
in the doping region from 11 - 18 \% in YBCO\cite{daou10} 
showed a strong $xy$ anisotropy. It sets in near the temperature where
the pseudogap appears so that the pseudogap region is interpreted as
the region with a finite $d$FSD in agreement with a theoretical study.\cite{hackl09} 
However, one should note that the regions where the in-plane anisotropy
has been observed by neutron scattering\cite{hinkov04,hinkov07,hinkov08} 
and by transport\cite{daou10} differ from each other so that it is 
difficult at present to reach clear-cut conclusions.
The experimental evidence for nematic order in cuprates has recently been critically
reviewed in Ref.~\onlinecite{singh10}.

Usually an emergent instability can be studied by measuring
the enhancement of the corresponding susceptibility. 
The susceptibility describing the $d$FSD is the $d$-wave charge 
compressibility,\cite{yamase00,metzner00,yamase05} which can be 
measured directly by Raman scattering. 
Hence Raman scattering  
can provide decisive evidence for a $d$FSD instability 
and its correlations in actual systems. 
However, despite various experimental studies in Sr327 and YBCO, 
Raman scattering experiments have not been 
reported to confirm a $d$FSD in those materials.

In this paper 
we provide theoretical predictions of the Raman scattering intensity from electrons 
and phonons near the $d$FSD instability in both normal and superconducting states  
by employing parameters appropriate to cuprate superconductors. 
In the superconducting state the Raman scattering intensity 
can be computed in terms of the non-interacting electron propagator, i.e., without considering
the damping of electrons. In the normal state, however, 
it is crucially important to include the electronic self-energy. 
We therefore include the Fock diagram for the self-energy, express it
in terms of the bosonic spectral function $\alpha^{2}F(\omega)$ and fit the latter to
the self-energy measured in ARPES.\cite{johnson01,zhao10} 

The paper is structured as follows. 
In Sec.~II we present formulas for electronic and phononic Raman scattering 
near a $d$FSD instability. Since the order parameter for $d$FSD fluctuations has
B$_{\rm 1g}$ symmetry for a square lattice  
only the B$_{\rm 1g}$ component of the Raman tensor and  B$_{\rm 1g}$ phonons will be
directly affected by order parameter fluctuations.
In Sec.~III we study Raman scattering for two different ways to  approach the $d$FSD instability.
In the first case the system is always in the normal state and the temperature is lowered
for a fixed doping in the underdoped region. In the second case 
we assume that at low temperature the $d$FSD instability lies in the superconducting
state and is reached by decreasing the doping. Results for both cases are given
in this section. Sec.~IV contains a detailed discussion of these results and our
conclusions.

\section{formalism}
In the following we will consider fermions on a square lattice which has the tetragonal point group
symmetry $D_{4h}$. Since the order parameter of the $d$FSD is the charge density 
with internal $d$-wave symmetry and zero total momentum, $d$FSD 
fluctuations will be most easily detected in the  B$_{\rm 1g}$ component of Raman scattering and
for a zone-center phonon with  B$_{\rm 1g}$ symmetry. We therefore will focus on these two
quantities in the following. Throughout the paper we will also use the lattice constant
of the square lattice as the length unit.

\subsection{Electronic Raman scattering}     
\begin{figure}[t]
\begin{center}
\includegraphics[width=8.5cm]{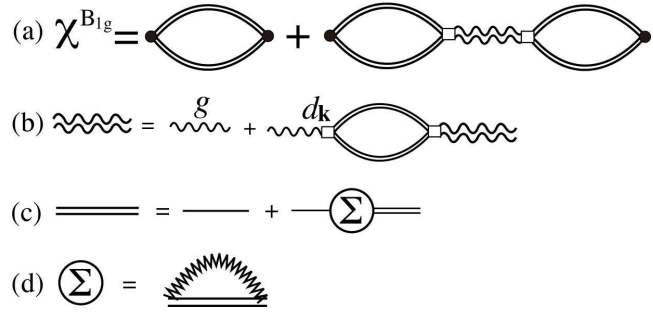}
\caption{(a) Graphical representation of $\chi^{\rm B_{1g}}$. 
The vertex with a circle (square) indicates the form factor $\gamma_{\vk}^{\rm B_{1g}}$ ($d_{\vk}$). 
(b) Effective electron-electron interaction driving the $d$FSD instability. 
(c) Full electronic Green's function. 
The single solid line denotes the free electron propagator with the dispersion $\epsilon_{\vk}$. 
(d) Electronic self-energy originating from the coupling to some bosonic fluctuations
represented by the sawlike line. } 
\label{raman-diagram}
\end{center}
\end{figure}

The electronic contribution to the  B$_{\rm 1g}$ Raman vertex is given in the 
effective mass approximation\cite{devereaux07}  
\be
\gamma_{\vk}^{\rm B_{1g}}=\frac{1}{2} \left(\frac{\partial^{2} \epsilon_{\vk}}{\partial k_{x}^{2}} 
- \frac{\partial^{2} \epsilon_{\vk}}{\partial k_{y}^{2}}\right)\, ,
\label{gamma}
\ee
where $\epsilon_{\vk}$ is the electronic dispersion, 
\be
\epsilon_{\vk} = -2 t (\cos k_x + \cos k_y) -4t' \cos k_x \cos k_y - 2t'' (\cos 2k_x + \cos 2k_y)\, ,
\label{dispersion}
\ee
with $t$, $t'$, and $t''$ being the nearest, second-nearest, and third-nearest neighbor hopping 
integrals, respectively. Inserting \eq{dispersion} into \eq{gamma} yields   
\be
\gamma_{\vk}^{\rm B_{1g}}= t (\cos k_x - \cos k_y) \left[1+ 8 t'' (\cos k_x + \cos k_y)/t \right]\,.
\label{B1g-vertex}
\ee 
The Raman scattering intensity $S(\omega)$ is given by 
\be
S(\omega) = - \frac{1}{\pi} [1 + b(\omega)] {\rm Im}\chi^{\rm B_{1g}} (\omega) \, ,
\label{Sw}
\ee
where $b(\omega)$ is the Bose function given 
by $(e^{\beta \omega}-1)^{-1}$ and $\beta^{-1} = T$ is the temperature. 
The quantity $\chi^{\rm B_{1g}}(\omega)$ is the retarded Green's function with two 
Raman vertices as end points and is given by 
\be
\chi^{\rm B_{1g}}(\omega) = - \frac{i} {N} \int_{0}^{\infty} {\rm d}t {\rm e}^{{\rm i} (\omega + 
{\rm i} 0^{+})t} \bra [\rho_{d}(t), \rho_{d}(0)] \ket \,,
\ee
where $N$ is the total number of lattice sites, $0^{+}$ an infinitesimally small quantity; 
$\bra \cdots \ket$ denotes the equilibrium expectation value, $[\cdot, \cdot]$ is the commutator, 
and $\rho_{d}(t)$ is the Heisenberg representation of the $d$-wave charge 
density operator 
\be
\rho_{d} = \sum_{\vk,\sigma} \gamma_{\vk}^{\rm B_{1g}} c^{\dagger}_{\vk \sigma} c_{\vk \sigma} 
\ee
with  $c^{\dagger}_{\vk \sigma} (c_{\vk \sigma})$ being the creation (annihilation) 
operator of electrons with spin $\sigma$ and momentum $\vk$. 
Within the RPA   $\chi^{\rm B_{1g}}(\omega)$ is
given by the bubble diagrams shown in Figs.~\ref{raman-diagram}(a) and (b). 
The double line represents
the electronic Green's function. In the normal state, which we discuss first,
self-energy corrections in the electronic Green's functions must be taken 
into account, as shown diagrammatically in \fig{raman-diagram}(c).  
Otherwise each bubble would become zero in the zero momentum limit
at every finite frequency. This means that mainly the
incoherent part of the Green's function contributes to Raman scattering in the
normal state. 

\begin{figure}[ht!]
\begin{center}
\includegraphics[width=9.5cm]{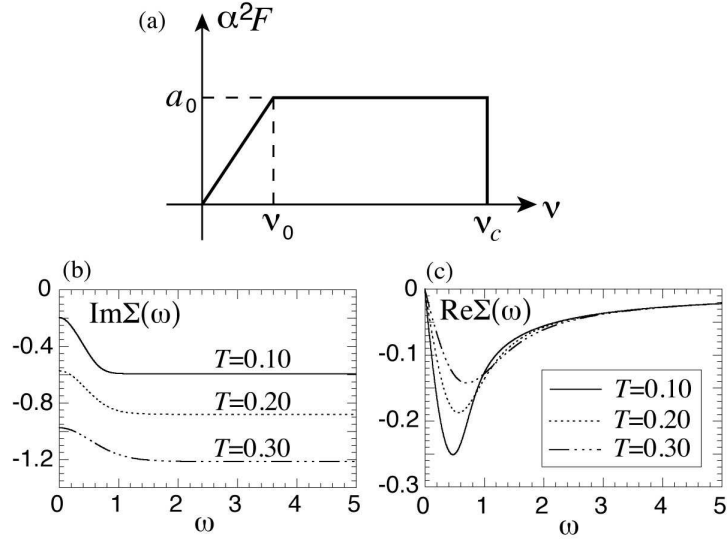}
\caption{(a) Model of $\alpha^{2}F(\nu)$. (b) Imaginary part and (c) real part 
of the electronic self-energy for several choices of $T$ in the normal state 
for $\nu_{0}=\nu_{c}/4$, $\nu_{c}=2/3$, and $a_{0}=1/4$. 
The energy unit is taken as $t$. 
} 
\label{alpha2F}
\end{center}
\end{figure}

In order to get a finite self-energy we consider the coupling of electrons  
to some bosonic fluctuations, described by the Fock diagram shown in \fig{raman-diagram}(d). 
Analytically one obtains, adopting the usual approximations in evaluating the
Eliashberg equations,\cite{mahan90}  
\be
{\rm Im}\Sigma(\omega) = - \pi \int_{0}^{\infty} {\rm d} \nu \,  \alpha^{2}F(\nu) \left[ 2 b(\nu) 
+ f (\nu-\omega) + f(\nu + \omega) \right] \,, 
\label{Imself-energy}
\ee
where $f(\nu) = ({\rm e}^{\beta \nu} +1 )^{-1}$ 
is the Fermi function, and
$\alpha^{2}F(\nu)$ specifies the bosonic spectral function.  
We have neglected the momentum dependence of $\alpha^{2}F$ for simplicity.  
Note that although the notation of $\alpha^{2}F(\nu)$ is often used in the 
context of a phonon spectrum, the bosonic modes in \fig{raman-diagram}(d) are arbitrary  
in our model. 
We model the function $\alpha^{2}F(\nu)$ with three parameters, 
$a_{0}$, $\nu_{0}$, and $\nu_c$, as illustrated in \fig{alpha2F}(a). Comparing with   
ARPES measurements in the normal state in cuprates\cite{johnson01,zhao10} 
we choose $\nu_{0}=\nu_c/4$, $\nu_c = 2t/3$, and $a_{0}=1/4$  
with $t \approx 150$ meV. 
In \fig{alpha2F}(b) we show Im$\Sigma (\omega)$ for several choices of $T$. 
The magnitude of Im$\Sigma (\omega)$ becomes larger with increasing $T$, indicating 
reductions of the lifetime of quasi-particles at higher $T$. 
As a function of energy, on the other hand, 
the longest lifetime of quasi-particle is realized on the Fermi surface, 
namely at $\omega=0$; the magnitude of Im$\Sigma(\omega)$ is  
enhanced with increasing $\omega$ and saturates to the value, 
$- \pi \int_{0}^{\infty} {\rm d}\nu\, \alpha^{2}F(\nu) [2b(\nu)+1]$ for $\omega \rightarrow \infty$. 
The real part of $\Sigma (\omega)$ is computed numerically 
from the Kramers-Kronig relation, 
${\rm Re} \Sigma (\omega) = \frac{1}{\pi} {\rm P.V.} \int_{-\infty}^{\infty} {\rm d} \nu 
\frac{{\rm Im} \Sigma (\nu) }{\nu - \omega}$,  
where the integral is defined as the principle value denoted by "P.V.".  
The obtained Re$\Sigma (\omega)$ is shown in \fig{alpha2F}(c).
The real part of $\Sigma (\omega)$ vanishes at $\omega=0$. 
Its magnitude forms a peak around $\omega \approx 0.5 t$ and is suppressed 
at high $\omega$ with a tail characterized by $\omega^{-1}$. 
Our self-energy reproduces well the data\cite{johnson01,zhao10} 
extracted from ARPES measurements in cuprate superconductors. 
\begin{figure}[th!]
\begin{center}
\includegraphics[width=9.5cm]{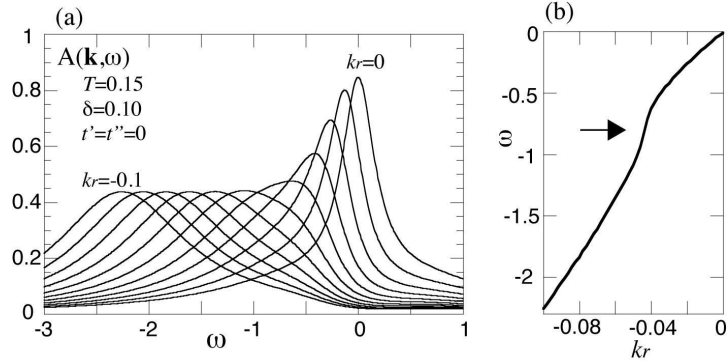}
\caption{(a) $A(\vk,\omega)$ as a function of $\omega$ at the momentum 
$\vk-\vk_{F}=2\pi k_r (1,1)$ with $k_r$ ranging from zero to $-0.10$ with an interval 
of 0.01; $\vk_{F}$ is the Fermi momentum along the $(0,0)-(\pi,\pi)$ direction. 
(b) The renormalized electronic dispersion. 
} 
\label{Akw}
\end{center}
\end{figure}
The spectral function of the full Green's function [\fig{raman-diagram}(c)] is given by 
\bea
&A(\vk,\omega) &= - \frac{1}{\pi} {\rm Im}G(\vk,\omega) 
\label{A-def}  \\ 
&&= - \frac{1}{\pi} \frac{{\rm Im} \Sigma (\omega)} {[\omega - (\epsilon_{\vk} -\mu) - {\rm Re} \Sigma (\omega)]^{2} + [{\rm Im}\Sigma(\omega)]^{2}} 
\label{A-def1}\,.
\eea
Here $\mu$ is the chemical potential which is approximately determined by the relation 
$\delta=1-\frac{2}{N}\sum_{\vk} f(\epsilon_{\vk}-\mu)$ 
for a given doping concentration $\delta$ and $T$. 
Figure~\ref{Akw}(a) shows $A(\vk,\omega)$ as a function of $\omega$ for several  
momenta along the $(0,0)-(\pi,\pi)$ direction. 
A relatively sharp peak is seen only close to the Fermi energy ($\omega=0$),
away from the Fermi surface it is substantially broadened 
because of the presence of the sizable Im$\Sigma(\omega)$. 
The peak position of $A(\vk,\omega)$ is plotted in \fig{Akw}(b). 
The renormalized electronic band dispersion displays a kink, as indicated by an arrow, 
due to the coupling to the bosonic fluctuations [\fig{raman-diagram}(d)]. 
These features in \fig{Akw} are qualitatively consistent with ARPES data.\cite{johnson01,zhao10}

A single bubble diagram in \fig{raman-diagram}(a) corresponds to the analytical expression, 
\be
\Pi^{\alpha \beta}(\omega) =\frac{2}{N} \sum_{\vk} \alpha_{\vk}\beta_{\vk} 
\int_{-\infty}^{\infty} {\rm d} \epsilon_{1} {\rm d} \epsilon_{2}  \, A(\vk,\epsilon_{1}) A(\vk,\epsilon_{2}) 
\frac{f(\epsilon_{1}) - f(\epsilon_{2})} {\epsilon_{1}+\omega-\epsilon_{2}+ {\rm i} 0^{+}} \,,
\label{Pi-ab}
\ee
where the form factors of the vertices are denoted by $\alpha_{\vk}$ and $\beta_{\vk}$. 
For the imaginary part Im$\Pi^{\alpha \beta}(\omega)$ we obtain  
\be
{\rm Im}\Pi^{\alpha \beta}(\omega) =\frac{2\pi}{N} \sum_{\vk} \alpha_{\vk}\beta_{\vk} 
\int_{-\infty}^{\infty} {\rm d} \epsilon \, A(\vk,\epsilon) A(\vk,\epsilon+\omega)  
\left[ f(\epsilon+\omega) - f(\epsilon) \right] \,.
\label{ImPi-ab}
\ee
The real part Re$\Pi^{\alpha \beta}(\omega)$ is determined from the Kramers-Kronig relation  
\be
{\rm Re} \Pi^{\alpha \beta} (\omega) = \frac{1}{\pi} {\rm P.V.} \int_{-\infty}^{\infty} {\rm d} \nu 
\frac{{\rm Im} \Pi^{\alpha \beta} (\nu) }{\nu - \omega}\, .
\label{RePi-ab}
\ee
Finally the Raman response function 
$\chi^{\rm B_{1g}}(\omega)$, described in \fig{raman-diagram}(a), is given by
\be
\chi^{\rm B_{1g}}(\omega)  = \Pi^{\gamma \gamma}(\omega) + 
\Pi^{\gamma d} (\omega) \frac{g}{1-g \Pi^{dd} (\omega)} \Pi^{d \gamma}(\omega)\,.
\label{XB1g}
\ee
The superscripts of $\Pi(\omega)$ "$\gamma$" and "$d$" indicate 
the form factors of the vertices of the bubble diagram, which are taken 
as $\gamma_{\vk}^{\rm B_{1g}}$ [\eq{B1g-vertex}] and $d_{\vk} =\cos k_{x} - \cos k_{y}$, respectively. 
The $d$-wave form factor comes from the electron-electron interaction which we write as  
\be
\frac{1}{2}\sum_{\vk \vk' \sigma \sigma'} g d_{\vk}d_{\vk'} c^{\dagger}_{\vk \sigma} c_{\vk \sigma} 
c^{\dagger}_{\vk' \sigma'} c_{\vk' \sigma'} \, ,
\label{fkk'} 
\ee
where $g(<0)$ is the coupling strength. 
This interaction generates the effective interaction shown in \fig{raman-diagram}(b) and  
drives the $d$FSD instability, as was extensively studied 
theoretically.\cite{yamase00,metzner00,kee0304,yamase05,lamas08}  
The condition for the instability is given by 
\be
1-g \Pi^{dd}(0)=0 \, .
\label{dFSDinstability}
\ee
From Eqs.~(\ref{A-def1}) and (\ref{ImPi-ab})-(\ref{XB1g}) 
we computed the Raman scattering intensity numerically employing 
the self-energy shown in \fig{alpha2F}. 

The selection of diagrams in \fig{raman-diagram} corresponds to the lowest-order
conserving approximation in the sense of Baym and Kadanoff.\cite{baym61}  
The diagrams shown in Fig.~3 in this reference also apply in
our case if we consider the dashed line as a sum of the interaction
of our \eq{fkk'} and our boson-mediated, retarded interaction which 
we have assumed to be independent of momentum.  The Hartree terms
to the self-energy can be omitted because they are either zero or
represent just a renormalization of the chemical potential.
The interaction \eq{fkk'} does not contribute in the thermodynamic limit
to the Fock term of the self-energy in contrast to the boson-mediated
interaction  which yielded the contribution given in \eq{Imself-energy}. 
The vertex is given as the functional derivative of the self-energy
with respect to the Green's function. Limiting ourselves to the
$d$-wave vertex 
we see that it is only the functional derivative of the Hartree term of the interaction \eq{fkk'} 
which contributes to the vertex and produces the chain of bubbles in the $d$-wave
susceptibility. This means that our approximation scheme respects
all conservation laws and should be free of artifacts due to an
inconsistent approximation.  

In the superconducting state the quasi-particle contribution to the Raman
scattering intensity is finite at finite frequencies. Since the self-energies
are also much smaller in the superconducting state compared to those in the normal state
it seems to be reasonable to neglect self-energy effects in this case.\cite{devereaux07} 
Assuming the following form for the $d$-wave superconducting gap  
\be
\Delta_{\vk} = \frac{1}{2} \Delta_{0} (\cos k_{x} - \cos k_{y})
\ee
and the band dispersion \eq{dispersion}, 
we obtain for the single bubble diagram [\fig{raman-diagram}(a) with $\Sigma=0$],
\be
\Pi^{\alpha \beta}(\omega) =\frac{1}{N} \sum_{\vk} \alpha_{\vk}\beta_{\vk} 
\frac{\Delta_{\vk}^{2}}{E_{\vk}^{2}} \tanh \frac{\beta E_{\vk}}{2} \left( 
\frac{1}{\omega - 2 E_{\vk} + {\rm i} \Gamma} - \frac{1}{\omega + 2 E_{\vk} + {\rm i} \Gamma} 
\right) \, ,
\label{Pisc-ab} 
\ee
where $E_{\vk} = \sqrt{(\epsilon_{\vk} - \mu)^{2} + \Delta_{\vk}^{2}}$. 
$\Gamma$ is an infinitesimally small  positive quantity which we approximate 
in our numerical calculations by $\Gamma=0.001t$. 
The chemical potential is approximately determined from the relation 
$\delta = \frac{1}{N} \sum_{\vk} \frac{\epsilon_{\vk}-\mu}{E_{\vk}} \tanh \frac{\beta E_{\vk}}{2}$. 
The Raman scattering intensity and response function are given again by the formulas  
Eqs.~(\ref{Sw}) and (\ref{XB1g}), respectively.

\subsection{Raman scattering from phonons}

\begin{figure}[t!]
\begin{center}
\includegraphics[width=8.5cm]{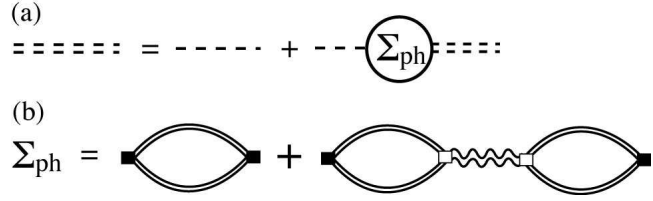}
\caption{Graphical representation of the phonon propagator (double dashed line); 
the single dashed line denotes the non-interacting phonon propagator, 
the solid square the form factor $g_{\rm ph} d_{\vk}$. The rest of the notation 
is the same as \fig{raman-diagram}. 
} 
\label{phonon-diagram}
\end{center}
\end{figure}

Raman scattering can also determine the spectral function of phonons. 
Since the interaction driving the $d$FSD instability couples to phonons 
with B$_{\rm 1g}$ symmetry, we focus on phonons with this symmetry. 
The corresponding electron-phonon coupling 
contains the $d$-wave form factor $d_{\bf k}$ and is given by 
\be
g_{\vk\vk} = g_{\rm ph} d_{\vk}, 
\label{electron-phonon}
\ee
where $g_{\vk \vk}$ is the electron-phonon 
matrix element for the electronic momentum $\vk$ and vanishing momentum for the phonon; 
$g_{\rm ph}$ is the coupling constant. 
The non-interacting retarded phonon propagator is given by 
\be
D_{0}(\omega) = \frac{1}{\omega - \omega_{0} + {\rm i}0^{+}} - 
 \frac{1}{\omega + \omega_{0} + {\rm i}0^{+} },
\ee
where $\omega_{0}$ is the energy of the zero-momentum B$_{\rm 1g}$ phonon, which 
corresponds to $\omega_{0}=4t/15 (\approx 40$ meV for $t\approx 150$ meV) 
in YBCO. \cite{pintschovius05} 
The full phonon propagator is given graphically in \fig{phonon-diagram}(a), namely, 
\be
D^{-1}(\omega) = D_{0}^{-1} (\omega) - \Sigma_{\rm ph}(\omega)\,.
\label{phonon-Green}
\ee
The free phonon propagator is renormalized by the electron-phonon interaction, 
which picks up the correlation function of the $d$FSD instability 
as shown in \fig{phonon-diagram}(b). 
The phonon self-energy $\Sigma_{\rm ph}(\omega)$ [\fig{phonon-diagram}(b)] has 
exactly the same structure as \fig{raman-diagram}(a) except for the difference of vertices. 
The computation of $\Sigma_{\rm ph}(\omega)$ is straightforward, yielding 
\bea
&\Sigma_{\rm ph}(\omega)&= g_{\rm ph}^{2} \frac{\Pi^{dd}(\omega)}{1-g \Pi^{dd}(\omega)} \\
&&=  g_{\rm ph}^{2} \widetilde{\Pi}^{dd}(\omega)
\label{self-ph}\,,
\eea
where we have introduced 
\be
\frac{1}{\widetilde{\Pi}^{dd}(\omega)}= \frac{1}{\Pi^{dd}(\omega)}-g \,.
\label{RPA-Pidd}
\ee
The quantity $\widetilde{\Pi}^{dd}(\omega)$ would become identical to 
$\chi^{\rm B_{1g}}(\omega)$ if "$\gamma$" were replaced by "$d$" in \eq{XB1g}.  
The Raman intensity $S_{\rm ph}$ 
for phonon scattering becomes
\be
S_{\rm ph}(\omega) = - \frac{1}{\pi} [ 1+b(\omega) ] {\rm Im} D(\omega) \, ,
\label{Sw-ph}
\ee
where from Eqs.~(\ref{phonon-Green}) and (\ref{self-ph}),  
\be
\frac{1}{\pi} {\rm Im}D(\omega) = \frac{4 \omega_{0}^{2} g_{\rm ph}^{2}}{\pi} 
\frac{{\rm Im} \widetilde{\Pi}^{dd} (\omega)} {
[\omega^{2} - \omega_{0}^{2} - 2 \omega_{0} g_{\rm ph}^{2} {\rm Re} 
\widetilde{\Pi}^{dd}(\omega)]^2 + 
[2 \omega_{0} g_{\rm ph}^{2} {\rm Im} \widetilde{\Pi}^{dd} (\omega)]^2} \, .
\label{ImD}
\ee
Since $\Pi^{dd}(\omega)$ has already been computed both in the normal and superconducting state 
in Eqs.~(\ref{ImPi-ab}), (\ref{RePi-ab}), and (\ref{Pisc-ab}), the Raman intensity 
$S_{\rm ph} (\omega)$ is easily obtained from Eqs.~(\ref{RPA-Pidd})-(\ref{ImD}).

\subsubsection{Renormalization of the $d$FSD by the electron-phonon coupling}
\begin{figure}[ht!]
\begin{center}
\includegraphics[width=8.5cm]{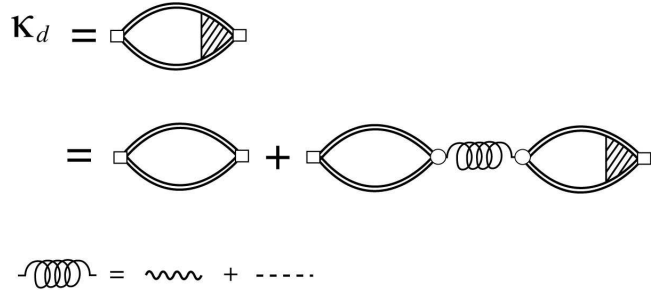}
\caption{Graphical representation of the $d$-wave charge compressibility. 
A spring denotes two interactions, 
the electron-electron interaction (wavy line) and 
the electron-phonon interaction (dashed line), and the corresponding form factors at the 
vertex (open circle) are  $d_{\vk}$ and $g_{\rm ph}d_{\vk}$, respectively. The shaded vertex
is defined by the equality of the first and second line, the 
rest of the notation is the same as in \fig{raman-diagram}.} 
\label{Kd-diagram}
\end{center}
\end{figure}

It is instructive to provide an expression of the static $d$-wave charge 
compressibility,\cite{yamase00,metzner00,yamase05} 
the susceptibility associated with the $d$FSD instability. 
This quantity is obtained by 
summing up the bubble diagrams connected by electron-electron and 
electron-phonon interactions as shown graphically in \fig{Kd-diagram}, 
that is, 
\bea
&\kappa_{d}&= - \frac{\Pi^{dd}(0)} {1-[g + g^{2}_{\rm ph} D_{0}(0)] \Pi^{dd}(0)} \\
\label{Kd}
&&=- \frac{\Pi^{dd}(0)} {1- \tilde{g} \Pi^{dd}(0)}, 
\label{Kd-2}
\eea
where 
\be
\tilde{g} = g - \frac{2 g_{\rm ph}^{2}}{\omega_{0}}
\label{tilde-g}
\ee
is a renormalized  coupling constant. 
Since both $g_{\rm ph}^{2}$ and $\omega_{0}$ are positive the original interaction $g (<0)$ 
is enhanced to become $|\tilde{g}| > |g|$. 
The coupling to the B$_{\rm 1g}$ phonon mode therefore increases the attractive interaction 
causing the $d$FSD instability by the amount $\frac{2g_{\rm ph}^{2}}{\omega_{0}}$. 
Therefore the $d$FSD instability can occur more easily and the instability condition 
\eq{dFSDinstability} is replaced by 
\be
1-\tilde{g} \Pi^{dd}(0)=0 \, .
\label{dFSDinstability-ph}
\ee

\subsubsection{Strength of the electron-phonon coupling}
An estimate for the coupling constant $g_{\rm ph}$ is obtained both from
first-principle calculations based on the local density approximation
(LDA) and from experiment. In general, the dimensionless coupling constant 
$\lambda$ for a phonon with energy $\omega_0$ and zero momentum is
defined by
\begin{equation}
\lambda = 2N(0) \bra |g_{{\bf k}{\bf k}}|^2 \ket_{\rm FS}/\omega_0,
\end{equation}
where $N(0)$ is the density of states at the Fermi energy for one
spin direction, $\bra \cdots \ket_{\rm FS}$ denotes an average over the Fermi surface, 
and $g_{{\bf k}{\bf k}}$ is defined by \eq{electron-phonon}. 
First-principle LDA calculations yielded  for the B$_{\rm 1g}$ phonon mode with 40 meV 
the values $\lambda = 0.02$\cite{thomsen90} and 0.06.\cite{bohnen03} 
On the other hand $\lambda$ is related to superconductivity-induced
self-energy effects of the phonon. It has been argued that the 
observed self-energy effects are compatible with these values for $\lambda$,
especially, with the first value.\cite{thomsen90}  A simple connection
between theory and experiment can be also obtained by noting that   
a phonon well below the superconducting gap at $T=0$ should
show a relative frequency softening of about 
$\delta \omega_0/\omega_0 =\lambda$.\cite{thomsen90}  
The above 40 meV phonon softens by $\sim 1$ meV due to
superconductivity\cite{krantz88,bakr09} yielding $\lambda \approx 0.02 - 0.03$ in rough agreement
with the theoretical prediction.

For a $B_{1g}$ phonon one has $g_{{\bf k}{\bf k}}=g_{\rm ph}d_{\bf k}$ 
and $\bra d_{\bf k}^{2} \ket_{\rm FS} = N_d(0)/N(0)$ where
$N_d(0)$ is the $d$-wave projected density at the Fermi energy, namely 
$N_{d}(0)=\int {\rm d}^{2}\vk \delta(\epsilon_{\vk}-\mu) d_{\vk}^{2}/(2\pi)^{2}$. 
We thus obtain
\begin{equation}
g_{\rm ph}^2 = \frac{\lambda \omega_0}{2N_d(0)}.
\end{equation}
Since $2N_d(0)$ is equal to the low-energy limit of a single
bubble at $T=0$, namely -$\Pi^{dd}(0)$,  we find $2N_d(0) \sim 1/|g|$, 
so that $g_{\rm ph}^2 \sim 0.4 \lambda t^{2}$ 
for $\omega_0 = 4t/15$ and $g=-1.5t$, yielding
values between 0.008 and 0.024 for $g_{\rm ph}^2$. In our numerical
calculations we use the representative value 0.02.

\section{Results}
Guided by experiments in  YBCO\cite{hinkov08,daou10} we would like to choose
one parameter set in our simple model such that the $d$FSD instability
is reached (a) with decreasing temperature at around $T \sim t/10$ 
in the normal state with a doping concentration $\delta =0.10$ 
and (b) with decreasing doping at around $\delta \sim 0.20$ for $T \approx 0$.
These conditions are approximately
fulfilled for  $t'=t''=0$ and $g=-1.5t$ in our model.\cite{misc-parameter} 
For convenience, we use $t$ as the energy unit in presenting our results. 
Experimentally, the effective $t$ has a value of about 150 meV.

\subsection{Electronic Raman scattering} 
\begin{figure*}[ht!]
\begin{center}
\includegraphics[width=11.5cm]{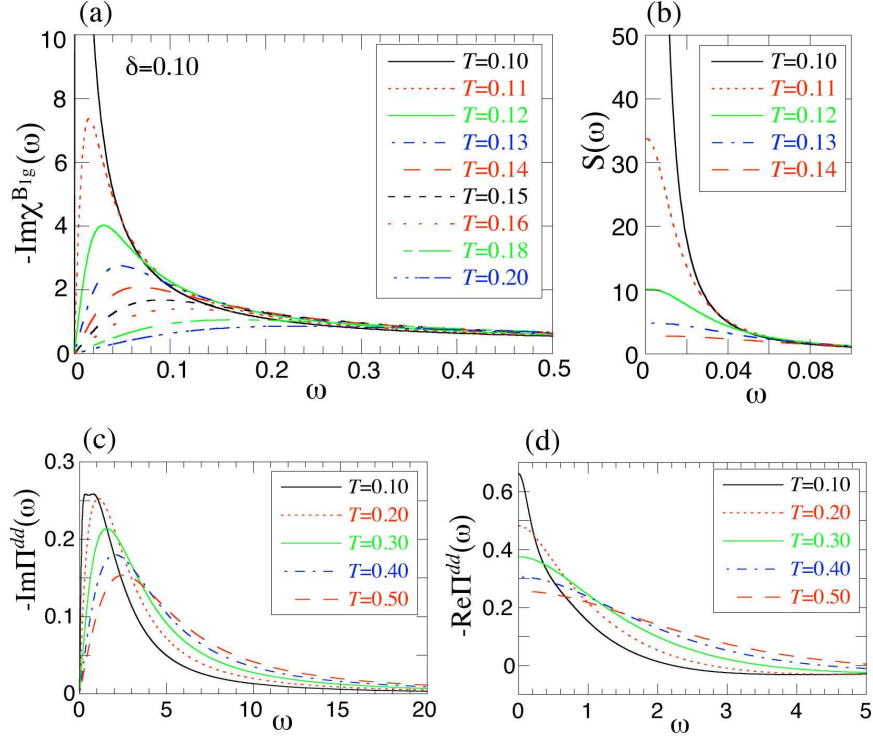}
\caption{(Color online) (a) $\omega$ dependence of 
Im$\chi^{\rm B_{1g}}(\omega)$ for a sequence of temperatures 
$T$ close to the $d$FSD instability at $T_{c}=0.098$ in the normal state 
at $\delta=0.10$. 
(b) Low energy region of $S(\omega)$ near the $d$FSD instability. 
(c) Im$\Pi^{dd}(\omega)$ and (d) Re$\Pi^{dd}(\omega)$ as a function of $\omega$ 
for several values of $T$. 
} 
\label{raman-normal}
\end{center}
\end{figure*}

In the normal state we fix the doping to $\delta=0.10$ and 
consider the temperature as a tuning parameter to approach the $d$FSD 
instability from high temperatures. For our parameters 
the $d$FSD instability occurs at $T=0.098$. 
In \fig{raman-normal}(a) we show the $\omega$ dependence of 
Im$\chi^{\rm B_{1g}}(\omega)$  
for a sequence of temperatures $T$ ranging from $0.10$ to $0.20$. 
At high $T$ the weight of Im$\chi^{\rm B_{1g}}(\omega)$ extends very broadly 
over the whole energy region shown in \fig{raman-normal}(a).  
With decreasing $T$ the low energy weight ($\omega <  0.2$) gradually 
increases and sharpens up to form a very steep peak near zero frequency. 
In \fig{raman-normal}(b) we plot the function $S(\omega)$,
defined in \eq{Sw}, 
which is measured in a Raman scattering experiment. 
Although the peak position is not exactly at $\omega=0$, 
$S(\omega)$ displays essentially a central peak already well away from the critical 
temperature. Its spectral weight increases strongly when the
critical temperature is approached from above. 
The energy dependence of Im$\Pi^{dd} (\omega)$ is shown in 
\fig{raman-normal}(c) for several values of $T$. 
While Im$\Pi^{dd}(\omega)$  exhibits also a pronounced peak  
its energy is much larger than that of Im$\chi^{\rm B_{1g}}(\omega)$. 
Moreover, the effect of temperature is much weaker in 
Im$\Pi^{dd}(\omega)$ than in Im$\chi^{\rm B_{1g}}(\omega)$. 
The real part of $\Pi^{dd}(\omega)$ 
is shown in \fig{raman-normal}(d). 
Its magnitude forms a broad peak at $\omega=0$ at high temperatures 
which sharpens up with decreasing temperature. 
Since the $d$FSD instability occurs when \eq{dFSDinstability}  
is fulfilled, collective fluctuations of the $d$FSD develop  
when the magnitude of Re$\Pi^{dd}(0)$ approaches $1/|g| =2/3$ 
with decreasing $T$.  Hence the very pronounced peak of 
Im$\chi^{\rm B_{1g}}(\omega)$ at low energy, seen in \fig{raman-normal}(a), 
is a direct consequence of the development of $d$FSD correlations. 

\begin{figure*}[th!]
\begin{center}
\includegraphics[width=6.5cm]{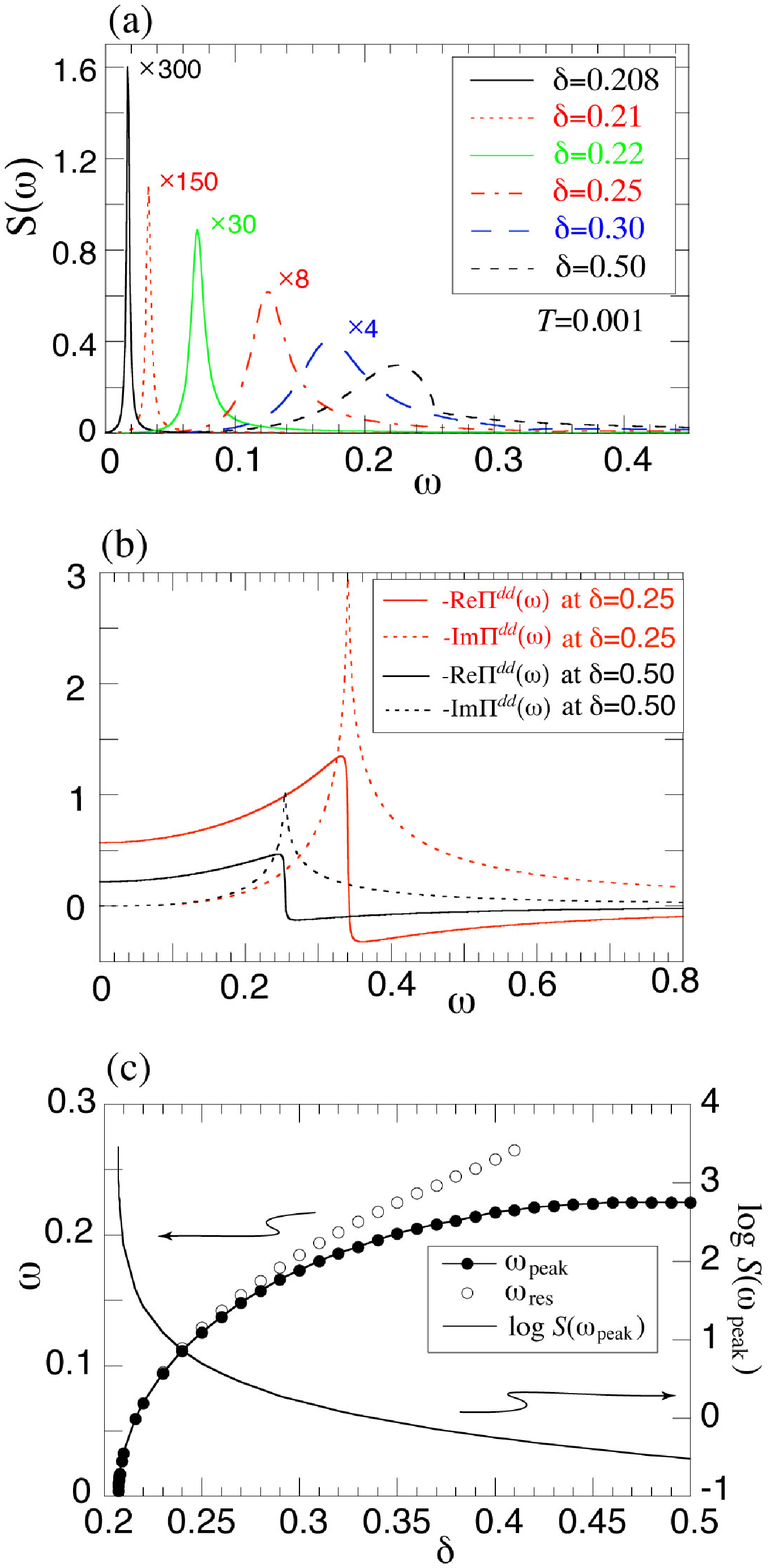}
\caption{(Color online) (a) $\omega$ dependence of $S(\omega)$ for a sequence of doping 
concentrations in the superconducting state at $T=0.001$; 
the actual value of $S(\omega)$ is obtained by multiplication with the factor  
indicated near each peak except for $\delta=0.50$. 
(b) $\omega$ dependence of Re$\Pi^{dd}(\omega)$ (solid line) and 
Im$\Pi^{dd}(\omega)$ (dashed line) at $\delta=0.25$ and $0.50$. 
(c) The peak position of $S(\omega)$  (solid circles) and its peak height (solid line) 
as a function of doping; also shown are the energies $\omega_{\rm res}$ (open circles);
the $d$FSD instability occurs at $\delta_{c}=0.207$. 
} 
\label{raman-sc}
\end{center}
\end{figure*}

Next we investigate the evolution of $d$FSD fluctuations 
in the superconducting state at $T=0.001$. Here we take the doping concentration $\delta$   
as a tuning parameter and approach the $d$FSD instability by decreasing $\delta$. 
We choose the superconducting gap amplitude to be $\Delta_{0}=t/5$ 
which seems to be reasonable for cuprate superconductors; 
the doping dependence of $\Delta_{0}$ is not important for our conclusions and  
will not be considered. 
We obtain $\delta_{c}=0.207$ for the critical doping rate where the $d$FSD instability 
occurs and consider the region $\delta > \delta_{c}$. 
Since there is no qualitative difference between $S(\omega)$ and 
Im$\chi^{\rm B_{1g}}(\omega)$ we present results only for $S(\omega)$. 
In \fig{raman-sc}(a) we show $S(\omega)$ as a function of $\omega$ for 
several choices of doping; similarly Im$\Pi^{dd}(\omega)$ 
and Re$\Pi^{dd}(\omega)$ are presented in \fig{raman-sc}(b) at 
$\delta=0.25$ and $0.50$. 
At $\delta=0.50$ the peak position of 
$S(\omega)$ is nearly the same as that of Im$\Pi^{dd}(\omega)$  
because the peak originates from individual excitations. 
Its position is determined approximately by 
$\omega = 2 |\Delta_{\vk}|$ at $\vk=\vk_{F}=(k_{F},0)$ or $(0,k_{F})$, 
where $k_{F}$ is the Fermi momentum along the $k_{x}$ or $k_y$ direction. 
With decreasing $\delta$ the peak of $S(\omega)$ shifts to lower 
energies, its half-width decreases and its height increases strongly.
The peak position substantially deviates from 
that of Im$\Pi^{dd}(\omega)$ [see the results at $\delta=0.25$ 
in Figs~\ref{raman-sc}(a) and (b)], indicating the development of collective 
fluctuations of the $d$FSD. 
In fact, the peak position of $S(\omega)$ is near the instability 
determined by the resonance condition, 
\be
1-g {\rm Re}\Pi^{dd} (\omega_{\rm res}) =0 \,. 
\label{resonance}
\ee
The resonance energy $\omega_{\rm res}$ is plotted in \fig{raman-sc}(c) 
together with the peak energy of $S(\omega)$ and its peak height.
At high $\delta$, \eq{resonance} does not have a solution and the peak of $S(\omega)$ 
must be attributed mainly to individual excitations. 
For $\delta \lesssim 0.40$ \eq{resonance} has a solution. 
It is seen that upon approaching $\delta_{c}$, $\omega_{\rm peak}$ becomes 
almost identical with $\omega_{\rm res}$. 
Since Im$\Pi^{dd}(\omega) \approx 0$ at $\omega \approx \omega_{\rm res}$  
the evolution of $S(\omega)$ in \fig{raman-sc}(a) indicates 
the development of a well-defined collective mode associated with the $d$FSD. 
Because of the collective fluctuations the peak intensity of $S(\omega)$ is strongly 
enhanced upon approaching $\delta_c$ and diverges at $\delta=\delta_c$. 
The peak energy vanishes as $\omega_{\rm peak} \sim (\delta-\delta_{c})^{1/2}$, 
which can be read off from \fig{raman-sc}(c). 

It is interesting to note the different evolution of $S(\omega)$ in 
the normal and the superconducting state. 
In the normal state the magnitude of Re$\Pi^{dd}(\omega)$ has a maximum at $\omega=0$ and 
decreases with $\omega$ [\fig{raman-normal}(d)], whereas in the superconducting state 
the magnitude of Re$\Pi^{dd}(0)$ corresponds to a local minimum and increases 
with $\omega$ [\fig{raman-sc}(b)]. In contrast to the superconducting case
the resonance condition \eq{resonance} is not fulfilled in the normal state
except at $T=T_{c}$ and $\omega=0$. 
This explains why $S(\omega)$ develops a central peak in the normal 
[\fig{raman-normal}(a)] and a soft mode in the superconducting [\fig{raman-sc}(a)]
state and why the width of the peaks is much larger in the normal than in the
superconducting state.

\subsection{Raman scattering from phonons} 
Raman scattering from ${\rm B_{1g}}$ phonons exhibits characteristic features near 
the $d$FSD instability. As a prominent example we consider the 40 meV phonon
in YBCO\cite{pintschovius05} which has in an approximate tetragonal classification,
where the chains are neglected,  B$_{\rm 1g}$ symmetry. Our parameter values become
$\omega_{0}=4/15$  and $g_{\rm ph}^{2} = 0.02$, as discussed in Sec.~II.B.2.
\begin{figure*}[ht!]
\begin{center}
\includegraphics[width=6.5cm]{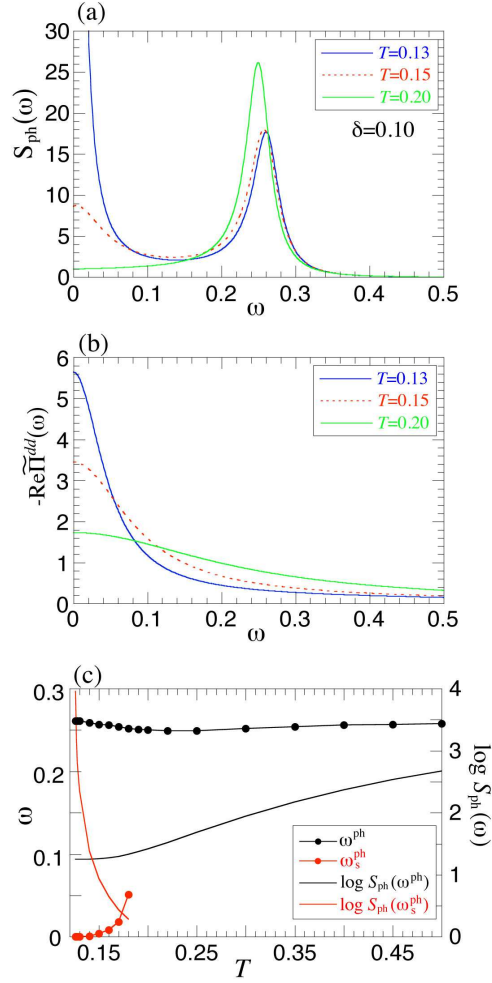}
\caption{(Color online) 
$\omega$ dependence of $S_{\rm ph}$ (a) and Re$\widetilde{\Pi}^{dd}(\omega)$ (b) 
for several choices of temperatures in the normal state at $\delta=0.10$; 
the renormalized critical temperature of the $d$FSD instability is $T_{c}=0.126$. 
(c) Peak positions $\omega^{\rm ph}$ and $\omega^{\rm ph}_{s} (< \omega^{\rm ph})$ 
 of $S_{\rm ph}(\omega)$ 
and their peak heights as a function of $T$. 
} 
\label{phonon-normal}
\end{center}
\end{figure*}

Figure~\ref{phonon-normal}(a) shows $S_{\rm ph}(\omega)$ in the normal state for 
$\delta=0.10$ and several values for $T$. In the presence of the electron-phonon interaction
the critical temperature occurs at $T_c = 0.126$ which is  
higher than in the case without electron-phonon interaction.
Well above this temperature, for instance, at $T=0.20$ $S_{\rm ph} (\omega)$ consists
of one single peak at  $\omega \approx \omega_{0}$ representing a usual quasi-harmonic
phonon. The peak position moves only slightly upwards, its spectral weight decreases
somewhat with decreasing temperature. However, at low frequencies dramatic changes
occur: approaching $T_c$ from high temperatures a central peak
develops. It extends over a rather broad energy region $\omega \lesssim 0.1$, but 
nearer to the instability its half-width decreases and its spectral weight increases
strongly.
It is caused by the coupling of the phonon to
$d$FSD fluctuations. The occurrence of a double-peak in the
phonon spectral function can be understood by studying the denominator of
the phonon spectral function \eq{ImD}, omitting Im$\widetilde{\Pi}^{dd}(\omega)$, 
\be
{\rm Res} (\omega) = \omega^{2}-\omega_{0}^{2} - 2 \omega_{0} g_{\rm ph}^{2} 
{\rm Re}\widetilde{\Pi}^{dd} (\omega)\,. 
\label{Res}
\ee
As shown in \fig{phonon-normal}(b), Re$\widetilde{\Pi}^{dd}(\omega)$ 
decreases monotonically with frequency 
and becomes small around $\omega \approx \omega_{0}$. 
Because $g_{\rm ph}^{2} =0.02$ is also small 
Res($\omega$) becomes zero at $\omega \approx \omega_{0}$ giving rise to 
the quasi-harmonic phonon mode. Since the magnitude of Re$\widetilde{\Pi}^{dd}(\omega)$ 
assumes its maximum at $\omega=0$  and increases there with decreasing $T$, 
it  eventually reaches the value 
$\omega_{0}/(2 g_{\rm ph}^{2})$ so that Res$(\omega)$ becomes zero also at $\omega=0$. 
This situation occurs just at $T_{c}$, 
because the expression Re$\widetilde{\Pi}^{dd}(0) = - \omega_{0}/(2 g_{\rm ph}^{2})$ reduces to 
Re$\Pi^{dd}(0) = 1/\tilde{g}$ via \eq{RPA-Pidd} which corresponds to the onset
of the $d$FSD instability [\eq{dFSDinstability-ph}].  
Hence both ${\rm Res}(\omega)$  and  Im$\widetilde{\Pi}^{dd}(\omega)$ for $\omega \approx 0$
are very small near $T_{c}$ which causes a central peak 
close to the $d$FSD instability.
The long tail in frequency of the central peak reflects the fact that 
the magnitudes of Res$(\omega)$ and Im$\widetilde{\Pi}^{dd}(\omega)$ only slowly 
increase with increasing $\omega$. 
In \fig{phonon-normal}(c) we plotted the peak positions and peak heights of
the phonon spectral function as a function of $T$. The lower peak position 
is denoted by $\omega_s^{\rm ph}$. 
We see that the central peak emerges well above $T_c =0.126$ and   
acquires rapidly a large spectral weight with decreasing $T$ which   
diverges at $T=T_{c}$. 
The high-frequency part of the phonon spectral function does not 
show a pronounced temperature dependence despite the proximity to the $d$FSD instability.  
The peak intensity near $\omega=\omega_{0}$ is suppressed at lower $T$ 
because of the increase of the magnitude of Im$\widetilde{\Pi}^{dd}(\omega)$ 
around $\omega \approx \omega_{0}$ upon approaching $T_{c}$.

\begin{figure*}[ht!]
\begin{center}
\includegraphics[width=6.5cm]{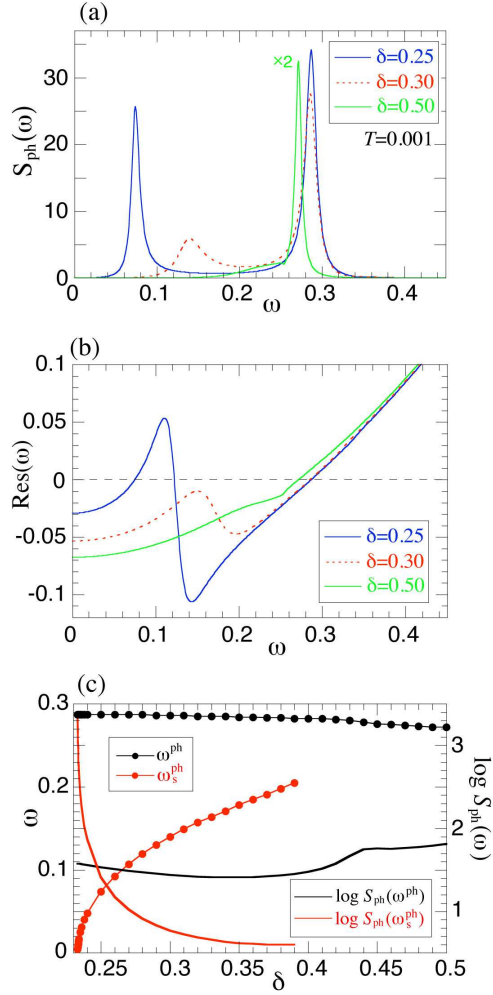}
\caption{(Color online) 
(a) $\omega$ dependence of $S_{\rm ph}(\omega)$ for several choices of $\delta$ 
in the superconducting state at $T=0.001$; the actual value at $\delta=0.50$ 
is obtained by multiplying with a factor of two. 
(b) $\omega$ dependence of Res$(\omega)$ for different $\delta$.
(c) Peak positions $\omega^{\rm ph}$ and $\omega^{\rm ph}_{s} (< \omega^{\rm ph})$  
of $S_{\rm ph}(\omega)$ as a function of $\delta$ together with their peak heights.
The $d$FSD instability occurs at $\delta_c=0.233$. 
} 
\label{phonon-sc}
\end{center}
\end{figure*}

The phonon spectral function in the superconducting state is shown in \fig{phonon-sc}(a) 
for several doping concentrations. At $\delta=0.50$ the spectral function shows a
quasi-harmonic phonon with one sharp peak at 
$\omega \fallingdotseq \omega_{1} \approx \omega_{0}$, 
where ${\rm Res} (\omega_{1}) = 0$ [\eq{Res} and \fig{phonon-sc}(b)].
With decreasing $\delta$ the position of this peak shifts only 
slightly to higher energies but is essentially unchanged. 
At $\delta=0.30$ an additional broad peak emerges at the low energy $\omega \approx 0.14$ 
in \fig{phonon-sc}(a).  
Its position is approximately given by the energy at which Res$(\omega)$ forms a 
local maximum, see  \fig{phonon-sc}(b). Decreasing $\delta$ further to 0.25
the lower peak becomes sharper, moves to lower energies and its spectral weight
increases. The equation Res$(\omega)=0$ has now three solutions, 
$\omega_{1}$, $\omega_{2}$, and $\omega_{3}$, see \fig{phonon-sc}(b), with 
$\omega_{1} > \omega_{2} > \omega_{3} \geq 0$. 
The solution $\omega_{1} \approx \omega_0$ yields the sharp high-frequency peak in $S_{\rm ph}(\omega)$, 
while the solution $\omega_{3}$ is responsible for the low-frequency peak. 
The solution of $\omega_{2}$ cannot produce a peak in $S_{\rm ph}(\omega)$ because 
Im$\widetilde{\Pi}^{dd}(\omega)$ has a peak near the energy $\omega_{2}$ and thus broadens 
out any structure in this frequency region.
These three solutions exist until $\omega_{3}$ becomes zero.
At this point the $d$FSD instability 
occurs which follows from a similar argument as given below \eq{Res}. 
In \fig{phonon-sc}{(c) we present the peak positions of $S_{\rm ph}(\omega)$ and 
their heights as a function of $\delta$. The high-frequency peak in the phonon spectral function,
appearing around 
$\omega \approx \omega_{0} =4/15$, displays only a very weak doping dependence 
in spite of the proximity to the $d$FSD instability. 
Its height also depends only weakly on $\delta$ on a logarithmic scale. 
The coupling of the phonon to collective fluctuations of the $d$FSD leads to the appearance 
of a second peak at low energy which softens in frequency and increases in intensity 
upon approaching the $d$FSD instability.
The lower peak energy vanishes as $\sim (\delta -\delta_{c})^{1/2}$ and its intensity diverges
when approaching the critical doping $\delta_{c} =0.233$. 
The emergent low-energy peak is a well-defined collective mode driven by 
fluctuations of the $d$FSD in the sense that the resonance condition 
Res$(\omega)\approx 0$ as well as Im$\widetilde{\Pi}^{dd}(\omega)\approx 0$  
is fulfilled at the peak energy.

It is intriguing to realize that the original quasi-harmonic B$_{1g}$ phonon mode does not 
behave like a soft-phonon when the $d$FSD instability is approached. Instead
the phonon spectral function splits into a high-frequency part which is
practically unaffected by the instability and an emergent low-frequency part which
behaves like a soft and a central mode in the superconducting and normal state, respectively. 
This double-peak structure in the spectral function is robust if the original 
phonon energy $\omega_{0}$ is sufficiently large. Otherwise, 
the phonon peak in the normal state may overlap 
with the emergent low-energy structure broadening 
the double-peak structure into a seemingly single peak. 
In the superconducting state $\omega_{0}$ should be chosen 
to be larger than the  peak energy in Im${\widetilde \Pi}^{dd}(\omega)$, 
which is approximately given by the peak position of $S(\omega)$ shown in \fig{raman-sc}(a). 
Otherwise, the original phonon mode softens 
down to zero energy upon approaching the $d$FSD instability 
and no additional low-energy peak emerges, 
in contrast to \fig{phonon-sc}(a).

\section{Discussion and conclusion}  

We have studied Raman scattering in a system where the interaction between electrons
drives the system towards a $d$FSD instability,
both in the normal and superconducting state. The electrons are
assumed to live on a square lattice with hopping amplitudes $t$, $t'$
and $t''$ between first, second and third nearest neighbors. The interaction
is a charge-density interaction with internal $d$-wave symmetry and interaction strength
$g$.  The parameters $t'$, $t''$, and $g$ are set up 
to mimic the strong tendency towards the $d$FSD instability
in YBCO. 
One could wonder whether our choice of $t'=t''=0$ and $g=-1.5t$ 
is unrealistic because the presence of substantial second- and third-nearest
neighbor hoppings is well known 
in cuprates\cite{tohyama00} and the value of $g$ simply seems too big. 
We would like to stress that the above parameter values 
should be interpreted as effective parameters within a 
phenomenological approach.\cite{misc-parameter}  
It is worth mentioning that a mean-field $d$FSD instability occurs in the
$t$-$J$ model with realistic parameters $t'$, $t''$, 
and $g (=-3J/8)$ for cuprates at lower carrier densities at 
temperatures as high as $0.2J$.\cite{yamase00} 
However, it is not easy to perform the above calculations directly for the $t$-$J$ model. 
Nevertheless, we believe that the essential features of such a more microscopic
approach are retained at least qualitatively in our simple phenomenological treatment. 

One result of our calculation is that the spectral
function of a B$_{\rm 1g}$ phonon exhibits a double-peak structure 
when the $d$FSD instability is approached 
as shown in Figs.~\ref{phonon-normal} and \ref{phonon-sc}. 
The double-peak structure might seem similar to the emergence 
of a central peak near structural phase transitions for several perovskites such as 
SrTiO$_{3}$,\cite{riste71,shapiro72} LaAlO$_{3}$,\cite{kjems73} and KMnF$_{3}$.\cite{gesi72} 
However the quasi-harmonic phonon also exhibits softening for these materials, 
in contrast to our results. Moreover the central mode in the experiments has been 
explained in terms of impurity scattering,\cite{halperin76}  
a different mechanism from ours. 
The double-peak structure we have obtained in Figs.~\ref{phonon-normal} and \ref{phonon-sc}  
can be interpreted as a general aspect in a coupled system of phonons and 
order parameter fluctuations. In fact, similar results to ours were obtained 
in a different context, for example, 
in pseudospin-phonon systems\cite{yamada74} and 
in superconductors with a strong electron-phonon 
coupling\cite{zeyher90,zeyher03} explaining  
the double peaks of Raman spectrum with E$_{2g}$ symmetry 
observed in MgB$_2$.\cite{quilty02}

For YBCO 
a strong tendency toward $xy$ symmetry breaking was 
observed.\cite{hinkov04,hinkov07,hinkov08,daou10} 
Its order parameter may be defined by 
\bea
&\phi&=\frac{1}{2} \sum_{\sigma} \bra c^{\dagger}_{i+x \sigma}c_{i \sigma}+
c^{\dagger}_{i \sigma}c_{i+x \sigma} \ket -
\bra c^{\dagger}_{i+y \sigma}c_{i \sigma}+
c^{\dagger}_{i \sigma}c_{i+y \sigma} \ket 
\label{phi1} \\
&&=\frac{1}{N}\sum_{\vk \sigma}d_{\vk}\bra c_{\vk}^{\dagger}c_{\vk \sigma} \ket \,,
\label{phi2}
\eea
where $i$ denotes the site on a square lattice 
and we have assumed that $\phi$ is constant. 
Equation~(\ref{phi1}) or (\ref{phi2}) is nothing but the 
order parameter of a $d$FSD instability.\cite{yamase00,metzner00,yamase05} 
It is characterized by Ising symmetry 
and thus two solutions, $\phi=\phi_{0}$ and $-\phi_{0}$, are degenerate. 
In order to favor either solution, it may be natural to apply a small external 
perturbation which breaks $xy$ symmetry in the CuO$_{2}$ plane. 
In fact the compound YBCO contains the CuO chains, which serves as 
a uniaxial strain. In this case, the $d$FSD instability becomes a crossover phenomenon, 
but the crossover is still sharp as far as the external anisotropy is weak which seems 
to hold in YBCO.

From the very strong anisotropy of the magnetic excitation spectrum in 
YBCO$_{6.45}$\cite{hinkov08} the presence of an underlying 
quantum critical point (QCP) has been conjectured in the doping range
$\delta \approx 8 - 10 \%.$\cite{kim08,huh08}  This conjecture could be   
tested in a rather direct way using Raman scattering in the superconducting state, 
see Figs.~\ref{raman-sc} and \ref{phonon-sc}. 
The measurement of the Nernst coefficient by Daou {\it et al.}\cite{daou10} 
determined the doping dependence of the $d$FSD instability in the region of
11 - 18 \% doping and suggested that the pseudogap temperature $T^{*}$ 
corresponds to the onset of the $d$FSD instability. 
Raman scattering can directly measure $d$FSD fluctuations and instabilities
generated by them, 
see Figs.~\ref{raman-normal} and \ref{phonon-normal}, and thus prove
the consistency of transport and light scattering data. 
Moreover, the resistivity measurement by Daou {\it et al.}\cite{daou09}  
suggested that $T^{*}$ goes down to zero in the overdoped region, implying the presence 
of a QCP associated with the $d$FSD instability inside the superconducting state.  
Theory\cite{kim08,huh08} and transport measurements,\cite{daou09,daou10} 
however, conjecture quite different values for the position of the QCP as a function of doping
which also could be clarified by Raman scattering in the superconducting state, 
see Figs.~\ref{raman-sc} and \ref{phonon-sc}. 
The neutron scattering experiments for YBCO$_{6,45}$,\cite{hinkov08} 
YBCO$_{6.6}$,\cite{hinkov04,hinkov07} 
and YBCO$_{6.85}$\cite{hinkov04} suggested a delicate interplay 
between the tendency towards a $d$FSD and the singlet pairing formation in agreement
with theory.\cite{yamase06,yamase09} Raman scattering  
around $T_{c}$ or the pseudogap temperature $T^{*}$ 
can directly reveal how the $d$FSD competes with the singlet pairing at different doping levels. 
Available Raman scattering data\cite{tassini05} for YBCO with 10 \% carrier doping 
do not suggest the strong enhancement of the low energy spectral weight, seen in 
\fig{raman-normal}, but the data were obtained only at a few temperatures. 
More detailed experimental studies including doping dependence are worth performing.

The La-based cuprate superconductors were extensively discussed
in terms of the charge-stripe order.\cite{kivelson03}  
However, the scenario based on the $d$FSD instability was also 
proposed.\cite{yamase00,yamase01,yamase07}  
Although the authors of Refs.~\onlinecite{tassini05} and \onlinecite{venturini02} 
interpreted the B$_{\rm 1g}$ Raman scattering spectra for La-based cuprates with 10 \% 
in terms of charge stripes, their data exhibit a spectrum very similar to 
\fig{raman-normal}(a), indicating direct evidence of the development of $d$FSD correlations. 
The data in Refs.~\onlinecite{tassini05} 
and \onlinecite{venturini02} are worth reconsidering. 

The $d$FSD is a generic feature in correlated electron systems 
and occurs in the $t$-$J$\cite{yamase00,edegger06,miyanaga06} and 
Hubbard\cite{metzner00,valenzuela01,wegner02,neumayr03} models, 
in systems where electrons interact 
via a central force\cite{quintanilla06,quintanilla08}, 
and quite generally in Fermi liquids with a van Hove saddle point.\cite{zverev10}  
Therefore the $d$FSD instability can be expected to occur 
in a variety of materials. 
In order to apply the present theory in the normal state 
we had to include self-energy effects. 
While quantitative features of the Raman spectrum certainly 
depend on details of the self-energy, 
it is not unreasonable to assume that its qualitative features associated with 
the proximity of the $d$FSD instability are rather robust. 
In this sense we hope that our results 
will serve to analyze Raman scattering data for various materials, which 
possibly lie close to a $d$FSD instability.  
In particular, compelling but indirect evidence for a $d$FSD instability 
has accumulated in Sr327 both experimentally\cite{grigera04,borzi07,rost09} 
and theoretically.\cite{kee05,doh07,yamase07b,yamase07c,puetter07,
yamase09b,ho08,fischer10,puetter10} 
It would be  
desirable to perform also Raman scattering measurements in this system 
to confirm the $d$FSD instability in a more direct and decisive way.

In summary, we have studied Raman scattering from electrons and phonons in the
normal and superconducting state near a $d$FSD instability. 
In the normal state  the inclusion of the electronic self-energy is vital 
for which we have used experimental input from
ARPES data in high-$T_c$ cuprates.\cite{johnson01,zhao10} Approaching the
$d$FSD instability from the normal state a central peak emerges both
in electronic scattering and in the spectral function of a phonon with $B_{\rm 1g}$
symmetry. Approaching the $d$FSD instability in the superconducting state
by decreasing doping concentrations 
a sharp soft mode appears in electronic Raman scattering.
This soft mode also appears in the spectral function of the phonon as an
additional low-energy feature whereas the usual phonon peak is nearly unaffected
by the proximity of the instability. Our study was motivated by recent 
transport measurements\cite{daou10} which suggest a $d$FSD instability in a wide
doping region in YBCO. Since Raman scattering measures directly the
correlation function of order parameter fluctuations associated with the
$d$FSD instability such measurements, together with our theoretical curves,
could confirm in a rather direct way the $d$FSD in real systems.

\begin{acknowledgments} 
We would like to thank  M. Bakr and B. Keimer for valuable discussions and 
W. Metzner for a critical reading of the manuscript. 
H.Y. appreciates warm hospitality of Max-Planck-Institute for Solid State Research and 
is partially supported by a Grant-in-Aid for  Scientific Research from Monkasho. 
\end{acknowledgments}


\bibliography{main.bib}

\begin{thebibliography}{71}
\bibitem[{mis()}]{misc-parameter}
\bibinfo{note}{We performed the same calculations for (i) $t'=-t/6$, $t''=0$,
  and $g=1.2t$ and (ii) $t'=-0.30t$, $t''=0.15t$, and $g=1.5t$, and confirmed
  the same conclusions.}
\end{thebibliography}

\end{document}